\documentclass[letterpaper]{article}

\usepackage{fontspec}

\usepackage{geometry}
\usepackage{setspace}

\usepackage[articletitle=false,maxauthors=0]{achemso}
\setkeys{acs}{doi = true}
\makeatletter
\DeclareRobustCommand\doi[1]{#1}
\makeatother

\usepackage{graphicx}
\usepackage{float}
\newfloat{scheme}{htbp}{los}
\floatname{scheme}{Scheme}
\floatname{chart}{Chart}
\newfloat{graph}{htbp}{loh}

\usepackage[dvipsnames]{xcolor}
\usepackage{chemformula} % Formulas using \ch{}
\usepackage[version = 4]{mhchem} % Formulas using \ce{}
\newcommand{\efgunit}{\ensuremath{\mathrm{V}\,\text{\AA}^{-2}}}
\providecommand{\doi}[1]{\href{https://doi.org/#1}{doi: #1}}
\usepackage{authblk}

\author[1]{Andrei Derevianko}
\author[1]{U. C. Perera}
\affil[1]{Department of Physics, University of Nevada, Reno, Nevada 89557, USA}
\author[2]{Marek Kro\'{s}nicki}
\author[2]{Kamil Nalikowski}
\affil[2]{Institute of Theoretical Physics and Astrophysics, University of Gda\'{n}sk, ul. Wita Stwosza 57, Gda\'{n}sk, 80-308, Poland}
\author[3]{H. W. T. Morgan}
\affil[3]{Department of Chemistry, University of Manchester, Oxford Road, Manchester M13 9PL, UK}
\author[4]{Valera Veryazov}
\affil[4]{Division of Computational Chemistry, Chemical Center, Lund University, P.O. Box 124, SE-221 00 Lund, Sweden}

\title{Electric Field Gradients in Multiconfigurational Embedded-Cluster Calculations of Solids
}
\date{*Email: valera.veryazov@teokem.lu.se}

\usepackage{subcaption}
\usepackage{threeparttable}

\newcommand{\thor}{\ensuremath{^{229}\textrm{Th}}}

\usepackage{siunitx}
\usepackage{graphicx}
\usepackage[colorlinks=true]{hyperref}
\usepackage{amsmath}
\usepackage{upgreek}
\usepackage{float}  
\usepackage{textcomp}
\usepackage{physics}

\usepackage{booktabs}

\usepackage{setspace}
\begin{document}

\maketitle

\begin{abstract}

Electric field gradients (EFGs) at nuclei contribute to atomic hyperfine structure and govern Nuclear Quadrupole Resonance (NQR) and M{\"o}ssbauer spectroscopy, providing a sensitive probe of local electronic structure and symmetry. EFGs can be obtained from electronic-structure calculations and inferred from spectroscopic measurements, linking theory and experiment.

In this work, we use representative molecular and crystalline examples to examine selected factors that control calculated EFGs, with particular emphasis on multiconfigurational embedded-cluster calculations for solids. We analyze the sensitivity of EFGs to geometry and to computational choices including basis set, correlation treatment, and active-space construction. We also discuss differences in EFG tensor conventions adopted by commonly used quantum-chemistry codes and their implications for comparing computational results. Periodic calculations are used where appropriate as reference data for assessing the embedded-cluster description.

Our results provide practical guidelines for the reliable computation and interpretation of EFGs as descriptors of the electronic structure and local chemical environment.

\end{abstract}

\section{Introduction}
\label{Sec:Introduction}

Electric field gradients (EFGs) quantify the inhomogeneity of the electric field at a nucleus arising from the surrounding electronic and nuclear charge distributions~\cite{Abragam1961PNM}. Their importance stems from the coupling to the nuclear quadrupole moment ($I \ge 1$), which leads to measurable splittings in hyperfine structure and in spectroscopies such as nuclear quadrupole resonance (NQR). The magnitude of this splitting is determined by the nuclear quadrupole coupling constant, which depends on the principal component $V_{zz}$ of the EFG tensor. EFGs therefore provide a direct link between experimental observables and the underlying electronic structure and chemical environment~\cite{Suits2006NQR,Wagle2024NQRSpectra,Errico2016CdMetalsEFG,Choudhary2020DFTEFG,Hartman2021FastEFG}.

While the EFG formalism is well established~\cite{Novick2011,Rinald2020,Autschbach2010,Lucken1972,Townes1949,Townes1952},
its application to multiconfigurational methods and embedded-cluster
calculations of solids has received considerably less attention.
Recent developments in embedded-cluster approaches now enable high-level
treatments of localized electronic states in solids, motivating an
assessment of EFGs within this framework.

In this work, we use representative examples to analyze selected factors
affecting EFG calculations in molecular and crystalline systems. Molecular
calculations are used to examine sign conventions, geometry dependence,
basis-set effects, and correlation treatment. For solids, the main focus is
on multiconfigurational embedded-cluster calculations and on whether the
embedding and active-space construction preserve the local symmetry required
for a physically meaningful EFG. Periodic DFT results are used as reference
values where appropriate rather than as a separate systematic benchmark of
periodic and cluster methods.

Because different quantum-chemistry packages adopt different conventions
for reporting the EFG tensor, direct comparison of published values may
be misleading. Throughout this work we use a single convention and
identify the corresponding conventions adopted by the computational
codes employed.

%A central issue addressed in this work is the inconsistency of EFG sign conventions across widely used quantum-chemistry packages. For example, \textsc{Gaussian}~\cite{Gaussian} and \textsc{MOLCAS}~\cite{oMolcas} report $V_{zz}$ with a sign opposite to that used in \textsc{ORCA}~\cite{ORCA} and \textsc{VASP}~\cite{VASP}. This ambiguity originates from the fact that zero--$B$-field NQR experiments are insensitive to the sign of $V_{zz}$~\cite{Drago}, while in other contexts the sign determines the ordering of nuclear-spin sublevels and the orientation of the principal axis~\cite{Pollack1997}. In this work, we reconcile these differences by adopting a consistent convention and providing a transparent connection between definitions.

This paper is organized as follows. Sec.~\ref{Sec:EFGs-HQ} introduces the EFG formalism and conventions. Sec.~\ref{Sec:ComputationalMethods} describes the computational methods. Sec.~\ref{Sec:discussion} presents the results, followed by conclusions. Unless stated otherwise, we use atomic units $|e|=\hbar=m_e\equiv 1$ and the Gaussian system of electromagnetic units, with $e>0$ denoting the elementary charge. One atomic unit of the EFG corresponds to $|e|/a_0^3 \approx $97.17~\efgunit.

\section{Electric field gradients and quadrupole interaction}
\label{Sec:EFGs-HQ}

\subsection{Model Hamiltonian}

The quadrupole interaction Hamiltonian $H_Q$ describes the interaction of the nuclear quadrupole moment $Q$ with the electric field gradient (EFG) generated by surrounding charges. In classical electrodynamics~\cite{JacksonEM}, it is written as
\begin{align}
H_Q = \frac{1}{6} \sum_{ij} Q_{ij} V_{ij} \,,
\label{Eq:HQ-Classical-EM}
\end{align}
where $V_{ij} \equiv \partial_i \partial_j V(\mathbf{r})$ is the EFG tensor evaluated at the nucleus.

Since $E_j = -\partial_j V$, the definition $V_{ij}=\partial_i \partial_j V$ is equivalent to $V_{ij}=-\partial_i E_j$. However, some literature and computational codes use the quantity $\partial_iE_j$ itself as the electric-field-gradient tensor. In this work we adopt $V_{ij} = \partial_i \partial_j V=-\partial_i E_j$. 
%These differences will be discussed in Sec.~\ref{Sec:EFGs-HQ:Conventions}.

The EFG tensor is real, symmetric, and traceless ($V_{xx}+V_{yy}+V_{zz}=0$). The electrostatic potential is given by
\begin{align}
 V(\mathbf{r}) = \sum_{B} \frac{Z_B}{|\mathbf{r} -\mathbf{R}_B| } 
 - \sum_k \frac{1}{|\mathbf{r} -\mathbf{r}_k| } \,,
\label{Eq:V_def}
\end{align}
which leads \cite{Olsen2002} to the EFG at nucleus $A$
\begin{align}
V_{ij}(\mathbf R_A)
= &\sum_{B\ne A} Z_B
\frac{3 R_{AB,i} R_{AB,j} - \delta_{ij} R_{AB}^{2}}{R_{AB}^{5}} \nonumber \\
&-
\mel{\Phi_{\mathrm{el}}}
{\sum_{k}
\frac{3 r_{Ak,i} r_{Ak,j} - \delta_{ij} r_{Ak}^{2}}{r_{Ak}^{5}}}
{\Phi_{\mathrm{el}}} \,.
\label{Eq:EFG_def}
\end{align}

Within a fixed nuclear spin $I$ manifold, the quadrupole tensor can be expressed in terms of spin operators as
\begin{align}
Q_{ij}=\frac{Q}{2\,I(2I-1)}
\Big(3I_i I_j+3I_j I_i-2\delta_{ij}\,I(I+1)\Big) \,.
\label{Eq:QijSpinOperators}
\end{align}

It is convenient to work in the principal axis frame (PAF), where the EFG tensor is diagonal. In this frame, the Hamiltonian becomes~\cite{Abragam1961}
\begin{equation}
H_Q = \frac{Q}{2I(2I - 1)} \left( V_{xx} I_x^2  + V_{yy} I_y^2 + V_{zz} I_z^2 \right) \,.
\label{Eq:HQ-xx-yy-zz}
\end{equation}

A commonly adopted convention is to order the principal components as
\begin{equation}
  |V_{zz}| \ge |V_{yy}| \ge |V_{xx}| \,,
\label{Eq:ViiSortingConvention}
\end{equation}
which fixes the labeling of the axes. This ordering is purely a convention: prior to its adoption, the identification of $x,y,z$ axes depended on the chosen molecular orientation, and different choices could lead to apparent sign changes of $V_{zz}$.

The asymmetry parameter is defined as
\begin{equation}
\eta \equiv (V_{xx} - V_{yy})/V_{zz} \,,
\label{Eq:etaDef}
\end{equation}
and, using the traceless condition, the Hamiltonian can be written in the standard form
\begin{equation}
H_Q = \frac{Q}{4I(2I - 1)}  V_{zz} \left[ 3I_z^2 - I(I + 1) + \eta \left( I_x^2 - I_y^2 \right) \right] \,.
\label{Eq:HQ-PAF-canonical}
\end{equation}

Spatial symmetry imposes strong constraints on the EFG tensor. For axially symmetric environments (e.g., tetragonal or hexagonal), $V_{xx}=V_{yy}=-\tfrac{1}{2}V_{zz}$ and $\eta=0$, so the tensor is fully determined by $V_{zz}$. In cubic symmetry, all tensor components vanish, $V_{xx}=V_{yy}=V_{zz}=0$, and no quadrupole splitting occurs.

The standard convention orders the principal components as
$|V_{zz}|\ge |V_{yy}|\ge |V_{xx}|$.
When $|V_{zz}|\approx|V_{yy}|$, as commonly occurs for $\eta\approx1$, the
assignment of the principal axes becomes numerically unstable.
Small structural perturbations or numerical uncertainties may then
interchange the two largest principal components, producing an apparent
change in the sign of $V_{zz}$ without any physical change in the system.

Therefore, care must be taken when interpreting $V_{zz}$ near
$\eta\approx1$. In such cases, it is preferable to analyze the full EFG
tensor or the resulting energy splittings, which are invariant with
respect to relabeling of the principal axes.

\subsection{EFG conventions}
\label{Sec:EFGs-HQ:Conventions}
\label{sec:symmetryefg}
There are several conventions for parametrizing the quadrupole interaction
Hamiltonian, which complicates comparisons between quantum-chemistry
codes and experimental data. Our self-contained derivation makes the
adopted definitions explicit and avoids ambiguities in interpreting the
reported EFGs.

Changing the sign convention for the EFG tensor does not affect the
asymmetry parameter $\eta$, but it reverses the ordering of the nuclear
sublevels split by the quadrupole interaction.

%Another ambiguity arises from the definition of the EFG itself. While we define $V_{ij}=\partial_i \partial_j V(\mathbf{r})$, a literal interpretation as the gradient of the electric field, $\partial_i E_j$, would introduce an overall sign change since $E_j = -\partial_j V(\mathbf{r})$. The asymmetry parameter $\eta$, Eq.~\eqref{Eq:etaDef}, is invariant under such sign changes; however, the overall sign of the quadrupole interaction is physically meaningful. Flipping the sign of $V_{zz}$ reverses the ordering of the nuclear energy levels within the EFG-split manifolds.

Some experimental observables are sensitive to the sign of $V_{zz}$,
whereas others are not. Conventional zero-field NQR experiments measure
transition frequencies within a manifold and therefore do not determine
the sign of $V_{zz}$~\cite{Smith71,Drago}. In contrast, hyperfine
structure measurements can determine the sign~\cite{WRJBook,BelDerJoh08-apd}.

The different conventions are reflected in the output of quantum-chemistry
codes. As discussed in Sec.~\ref{Sec:EFG-molecules},
\textsc{Gaussian}~\cite{Gaussian} and
\textsc{MOLCAS}~\cite{MOLCAS} employ the opposite sign convention from
\textsc{Dalton}~\cite{Dalton},
\textsc{ORCA}~\cite{ORCA}, and
\textsc{VASP}~\cite{VASP}. Throughout this work all EFGs are reported
using a single convention.

Finally, the nuclear quadrupole moment $Q$ may be defined either with or
without the elementary charge~\cite{Smith71,Pyykko2008}. In
Eq.~\eqref{Eq:HQ-Classical-EM}, $Q_{ij}$ is defined as a moment of the
nuclear charge density and therefore includes the charge by
construction. Consistent units ensure an unambiguous interpretation.

\section{Computational methods}
\label{Sec:ComputationalMethods}

We employed Hartree--Fock (HF)~\cite{Fock1930},
density-functional theory (DFT)~\cite{KSDFT},
coupled-cluster theory with single and double excitations (CCSD),
and multiconfigurational methods, specifically CASSCF and
CASPT2~\cite{CASSCF,Andersson_90,ThePinkBook}.
These approaches were used to assess the sensitivity of EFGs to the
level of electronic-structure treatment.

Solid-state EFGs were considered using embedded-cluster calculations and,
where available, periodic-boundary-condition (PBC) DFT reference results~\cite{Payne1992,VASP}.
Embedded-cluster methods~\cite{Catlow,Kaplan} treat a finite quantum-mechanical
region in an effective representation of the surrounding lattice. A finite
cluster does not reproduce the complete Bloch-state description of an infinite
solid; its applicability therefore depends on an adequate representation of the
local electronic structure and crystalline environment for the property of
interest. In the present work this is tested through symmetry requirements and
comparison with periodic or experimental reference data where available. The
embedded-cluster formulation is particularly useful here because it permits
high-level multiconfigurational and post-Hartree--Fock treatments relevant to
the spectroscopic applications motivating this work.

Embedded-cluster calculations were performed using \textsc{MOLCAS}~8.6~\cite{oMolcas}. The quantum cluster (QC) was treated at the multiconfigurational level and embedded in a shell of {\em ab initio} model potentials (AIMPs) and an extended point-charge environment generated with the SCEPIC procedure~\cite{LARSSON2022111549}. The embedding contains more than $5\times 10^{4}$ point charges. Their positions and values are generated algorithmically so as to reproduce the electrostatic potential in the quantum-cluster region while compensating the dipole and higher multipole moments through fourth order. Thus, the embedding is defined by the generation procedure rather than by a fixed manually selected number of charges. The construction and validation of the AIMP and point-charge environment are described in detail in Ref.~\cite{LARSSON2022111549}; here we focus on its consequences for the EFG and symmetry of the embedded quantum region.

EFGs were computed with several electronic-structure codes to verify the
sign conventions adopted in their reported EFG tensors. Program-appropriate numerical representations were used: localized orbital bases in \textsc{CRYSTAL}, a mixed Gaussian/plane-wave representation in \textsc{CP2K}, and a plane-wave representation in \textsc{VASP}. The basis quality was chosen to be comparable to the converged calculations discussed in the Supporting Information. Since the purpose of these cross-code calculations is to establish EFG conventions rather than to benchmark basis representations, no attempt was made to impose an identical basis representation across fundamentally different electronic-structure implementations.

In \textsc{MOLCAS}, ANO-RCC basis
sets~\cite{ALMLOF,ANO-RCC,Zobel2019} together with the second-order
Douglas--Kroll--Hess (DKH2) scalar-relativistic
Hamiltonian~\cite{Reiher2004_DKH} were employed. Details of the basis
sets and convergence tests are provided in the Supporting Information.
\section{Results and discussion}\label{Sec:discussion}

\subsection{EFG sign convention}
\label{Sec:EFG-molecules}

As discussed in Sec.~\ref{Sec:EFGs-HQ}, different quantum-chemistry codes employ different sign conventions for the EFG. Here we compare several widely used electronic-structure packages to identify the conventions adopted in their reported EFG tensors.

In this section, we focus on the water molecule $\ce{H2O}$~\cite{Verhoeven1969}
as a simple test system that can be treated with a variety of electronic-structure
methods and computational codes. We also examined several transition-metal
complexes and found the same program-dependent sign conventions for the EFG.
These additional calculations are therefore omitted for brevity.

%The calculations reported in Table~\ref{tab:efg_hf} were performed using the default settings of each electronic-structure package. The objective is not to benchmark the numerical agreement between codes, but to establish the sign convention adopted in their reported EFG tensors. Differences in the numerical values are therefore expected because of different basis sets, numerical thresholds, and implementation details.

\begin{table}[ht]
\centering
\caption{Principal component $V_{zz}$ (\efgunit) and asymmetry parameter $\eta$
for the $^{17}$O nucleus in H$_2$O.
Programs marked with an asterisk ($^\ast$) report the EFG tensor with the
opposite sign to the convention adopted in Eq.~(\ref{Eq:EFG_def}).
The calculations are intended to identify the sign convention used by each
program rather than to provide a quantitative benchmark. All calculations were
performed using the default settings of the corresponding program and the
program-appropriate numerical representations described in the Computational
Methods and Supporting Information. 
Consequently, the numerical magnitudes are not intended for code-to-code
benchmarking: they reflect differences in basis sets or basis representations,
default numerical thresholds, implementation details, and, between rows using
different electronic-structure methods, the treatment of electron correlation.
}
\label{tab:efg_hf}
\begin{tabular}{llrr}
\hline
Method & Software & $V_{zz}$ & $\eta$ \\
\hline
HF &
MOLCAS$^\ast$   & 181  & 0.79 \\
&
Gaussian$^\ast$ & 173  & 0.79 \\
&
CRYSTAL         & -182 & 0.80 \\
&
ORCA            & -182 & 0.79 \\
&
VASP            & -212 & 0.75 \\
\hline
DFT--PBE &
MOLCAS$^\ast$   & 169  & 0.74 \\
&
Gaussian$^\ast$ & 162  & 0.75 \\
&
CP2K$^\ast$     & 142  & 0.70 \\
&
ORCA            & -176 & 0.75 \\
&
VASP            & -195 & 0.71 \\
\hline
CCSD &
ORCA            & -169 & 0.76 \\
\hline
CASSCF &
MOLCAS$^\ast$   & 165  & 0.78 \\
\hline
CASPT2 &
MOLCAS$^\ast$   & 165  & 0.77 \\
\hline
Experiment$^{a}$ & & -164(2) & 0.75(2) \\
\hline
\end{tabular}

\vspace{1mm}
\footnotesize
$^{a}$ Reference~\cite{Verhoeven1969}.
\end{table}

Table~\ref{tab:efg_hf} summarizes the computed principal EFG component $V_{zz}$ and asymmetry parameter $\eta$ for oxygen in water. Experimental EFG values, where available, are not measured directly but inferred from quadrupole coupling constants.

We compare results from several widely used electronic-structure codes
using the convention defined in Sec.~\ref{Sec:EFGs-HQ}.
\textsc{ORCA}, \textsc{CRYSTAL}, and \textsc{VASP} follow this
convention, whereas \textsc{MOLCAS}, \textsc{CP2K}, and
\textsc{Gaussian} (marked by an asterisk in
Table~\ref{tab:efg_hf}) employ the opposite sign.
The values in Table~\ref{tab:efg_hf} are reported exactly as printed by
each program; throughout the remainder of this work all EFGs are
expressed using the \textsc{ORCA}/\textsc{VASP} convention.

The numerical magnitudes differ appreciably in some cases. These differences
should not be interpreted as a benchmark of the programs because the entries
were obtained with different basis sets or basis representations, numerical
settings, and, across methods, different treatments of electron correlation.
The purpose of Table~\ref{tab:efg_hf} is instead to identify the sign convention
reported by each program; that convention produces the clear systematic sign
difference in the table.

Having established the computational conventions with a molecular test case, we now turn to the main solid-state application of this work: multiconfigurational embedded-cluster calculations.

\subsection{EFG in embedded cluster calculations}\label{sec:EFG-clusters}

In the embedded-cluster framework, implemented in \textsc{MOLCAS}~\cite{LARSSON2022111549} , a finite region of the crystal (the quantum cluster, QC) is treated explicitly at the electronic-structure level, while the surrounding lattice is represented by an effective electrostatic environment. This environment is constructed from {\em ab initio} model potentials (AIMPs), which account for short-range Coulomb and exchange interactions, and a set of point charges that reproduce the long-range Madelung potential and compensate the formal charge of the QC. This partitioning enables a high-level treatment of localized electronic states while retaining the electrostatic influence of the crystal.

Unlike periodic calculations, where crystal symmetry is preserved by
construction, embedded-cluster models require explicit verification that
the cluster and embedding preserve the local symmetry. The EFG at the
central atom contains both electronic contributions from the quantum
cluster and nuclear contributions from the surrounding embedding
[Eq.~\eqref{Eq:EFG_def}], making it particularly sensitive to
imperfections in the cluster construction and embedding.

We selected $^{43}$Ca in \ce{CaF2} and $^{59}$Co in \ce{LaCoO3} as
benchmark systems. Cubic \ce{CaF2} provides a stringent test because
symmetry requires all EFG tensor components to vanish. We also consider
the substitutional defect \ce{Th}:\ce{CaF2}, where the charged impurity tests
whether the embedding preserves the local symmetry in the presence of
charge compensation.
%We selected $^{43}$Ca in \ce{CaF2} and $^{59}$Co in \ce{LaCoO3} as benchmark systems to test how well the embedding procedure preserves crystallographic symmetry. Calcium fluoride has a cubic structure and is computationally simple; its octahedral site symmetry implies that all components of the EFG tensor vanish ($\eta = 0$) at every nuclear site. This makes it a convenient reference system for assessing whether embedded-cluster calculations preserve symmetry and avoid spurious EFG contributions.
%In addition to pristine \ce{CaF2}, we consider a substitutional defect system, \ce{Th}:\ce{CaF2}, where tetravalent Th replaces divalent Ca. Although the host lattice is cubic, the charge imbalance introduced by the defect may perturb the local symmetry. In periodic calculations, this requires artificial charge compensation \cite{hiraki2025lasermossbauerspectroscopy229th}, whereas in the embedded-cluster approach the charge is balanced by the embedding. This system therefore provides a sensitive test of how well the embedding preserves the local symmetry in the presence of a charged defect.
Finally, we consider \ce{LaCoO3}, which has a rhombohedrally distorted perovskite structure. In this case, the reduced symmetry leads to a nonzero $V_{zz}$, providing a complementary example where the EFG reflects a genuine structural distortion rather than numerical artifacts.

\begin{table}[ht!]
\centering
\caption{Principal component $V_{zz}$ (in \efgunit) and asymmetry parameter $\eta$ calculated at different nuclear centers for quantum clusters embedded into the \ce{CaF2} and \ce{LaCoO3} lattices.
All embedded-cluster calculations were carried out with
\textsc{MOLCAS}. The reported $V_{ii}$ values have been converted to the
sign convention adopted in this work.
%In Supplementary we studied the change of the active space of Th cluster. 
} 
\label{tab:efg_cluster}
%\begin{ruledtabular}
\begin{tabular}{ccclr ll}
Nuclear Center &Chemical Formula &Quantum Cluster & Method & \multicolumn{1}{c}{$V_{zz}$} & \multicolumn{1}{c}{$\eta$} \\
\midrule
$^{43}$\textbf{Ca } & \ce{CaF2}
            &  \ce{\textbf{Ca}F8Ca12^{18+}} & HF & 0.0 & 0.00 \\
            &&                               & CASSCF (6e,12o)&  0.0 & 0.00 \\
            &&                               & CASPT2 (6e,12o)&  0.0 & 0.00\\
            &&                               & By symmetry & 0.0 & 0.00\\
\midrule
$^{229}$\textbf{Th} & \ce{Th{\rm :}(CaF2)}&\ce{\textbf{Th}F8Ca12^{20+}} & HF    & 0.0    & 0.00\\
&& & CASSCF (6e,16o)        &  0.0        & 0.00 \\
&& & CASPT2  (6e,16o)       &  0.0        & 0.00 \\
% && & CASSCF (12e,12o)      &  -5.9        & 0.02 \\
% && & CASPT2  (12e,12o)     &  -7.7        & 0.02 \\
                    &&   &  DFT–PBE                      & 0.0&0.00 \\  
                    && --  &  PBC-DFT–PBE                      & 0.0 &0.00  \\
                                && & By symmetry & 0.0 & 0.00\\
[4pt] \midrule
% $^{59}$\textbf{Co}    &\ce{LaCoO3}&\ce{\textbf{Co}O6^{9-}}   &   HF            & 0.0  & 0.00$^\dagger$  & \\
%                       && cubic cluster            &CASSCF (42e,23o) &  0.0  & 0.00$^\dagger$  & \\
%                       &&                          &CASPT2 (42e,23o) &  0.0  & 0.00$^\dagger$  & \\\cmidrule(lr){3-7}
%                       &&\ce{\textbf{Co}O6^{9-}}   &   HF            & 10.6  & 0.00  & \\
%                       &&tetragonal cluster        &CASSCF (42e,23o) & 10.6  & 0.00  & \\
%                       &&                          &CASPT2 (42e,23o) & 10.4  & 0.00  & \\\cmidrule(lr){3-7}
                       $^{59}$\textbf{Co}&\ce{LaCoO3}& \ce{\textbf{Co}O6^{9-}}  & HF               & -4.6  & 0.00  \\
                      &&                          &CASSCF  (42e,23o) & -7.4  & 0.00  \\
                      &&                          &CASPT2  (42e,23o) & -7.8  & 0.00  \\ %\cmidrule(lr){3-7}
                      &&       &LDA+U  & -8.8(6)$^a$ & \\
                      &&       &Experiment  & $\pm$ 8.1(8)$^b$  & 0.00\\

\end{tabular}
%\end{ruledtabular}
\begin{tablenotes}[flushleft]
        \small
        \item $^a$Reference \cite{Hsu2010}.
        \item $^b$Reference \cite{Itoh2013}.
\end{tablenotes}
\end{table}

The EFG calculations for the \ce{CaF8Ca12^{18+}} and
\ce{ThF8Ca12^{20+}} clusters are summarized in
Table~\ref{tab:efg_cluster}. Note that the formal charge of the quantum
cluster is fully compensated by the embedding environment, so that the
embedded system represents a charge-neutral crystal. In the notation used
here, the central atom is listed first, followed by the successive coordination
shells and the formal charge of the quantum cluster (QC).

For the cubic systems, all components of the EFG tensor should vanish by
symmetry. Recovering $V_{ii}=0$ therefore provides a stringent test of
both the embedding procedure and the multiconfigurational description.
The CASSCF calculations satisfy this condition, provided that the active
space includes all symmetry-equivalent crystal orbitals originating from
the open shells of the Th impurity.

The effects of lattice distortion in the embedded-cluster model are analyzed for \ce{LaCoO3}, a rhombohedrally distorted perovskite~\cite{Rao1970,Radaelli2002}. The rhombohedral structure may be viewed as a distortion of the cubic phase along the body diagonal ($C_{3}$ axis). In the cubic limit, the EFG at the Co site vanishes by symmetry, providing a natural reference for assessing the effect of structural distortions.

For the experimental rhombohedral structure and the symmetrized cubic reference, the embedding was constructed using the same AIMPs and point charges, so that changes in the calculated EFG can be associated with the imposed structural distortion. Table~\ref{tab:efg_cluster} reports the values for the rhombohedral structure, whereas the cubic reference and controlled distortions about it are shown in Fig.~\ref{fig:tet-dist}. In the cubic limit the EFG vanishes by symmetry; away from that limit the magnitude of $V_{zz}$ depends both on the geometry and on the electronic-structure treatment.

Experimentally, $V_{zz}$ for \ce{LaCoO3} is inferred from NMR measurements of the quadrupole coupling constant~\cite{Itoh2013}. Because the observable depends on $V_{zz}^{2}$, the sign of $V_{zz}$ cannot be determined, yielding $V_{zz}=\pm 8.1(8)\,
$\efgunit~\cite{Pyykko2008}. 
The CASSCF and CASPT2 magnitudes are close to this experimental estimate, whereas the HF value is substantially smaller, illustrating that the magnitude of a symmetry-allowed EFG can depend appreciably on the electronic-structure treatment.

The agreement of a particular DFT result with experiment in this example should not be interpreted as establishing a general hierarchy of EFG accuracy between DFT and multiconfigurational wave-function methods. The present calculations use representative systems to examine sensitivity and symmetry rather than a common benchmark protocol for method accuracy or computational cost. In particular, the multiconfigurational embedded-cluster treatment is motivated by applications in which excited states and correlated spectroscopy require a wave-function description beyond the scope of the present periodic reference calculations.

These results show that, for the systems tested here, a consistently
constructed quantum cluster and embedding preserve the symmetry constraints on
the EFG. Symmetry determines whether the EFG must vanish, while the magnitude
of a symmetry-allowed nonzero $V_{zz}$ remains sensitive to the
 electronic-structure treatment, as illustrated by the HF, CASSCF, and CASPT2
values for \ce{LaCoO3}.

\subsection{EFG sensitivity to geometry change\label{sec:geometry_sensitivity} }

\begin{figure}[ht!]
    \centering 
    
{\includegraphics[width=0.5\textwidth]{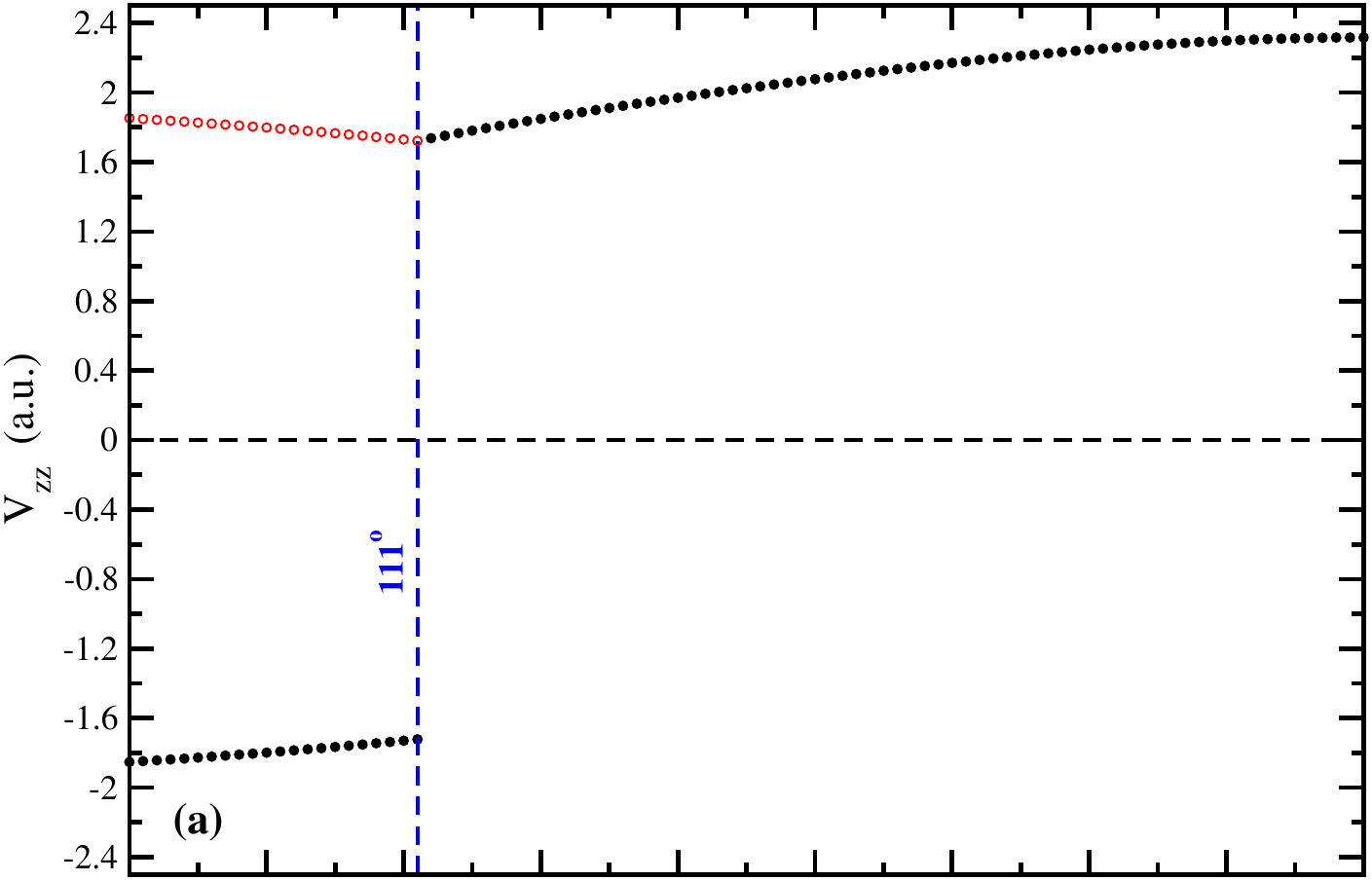}}

{\includegraphics[width=0.5\textwidth]{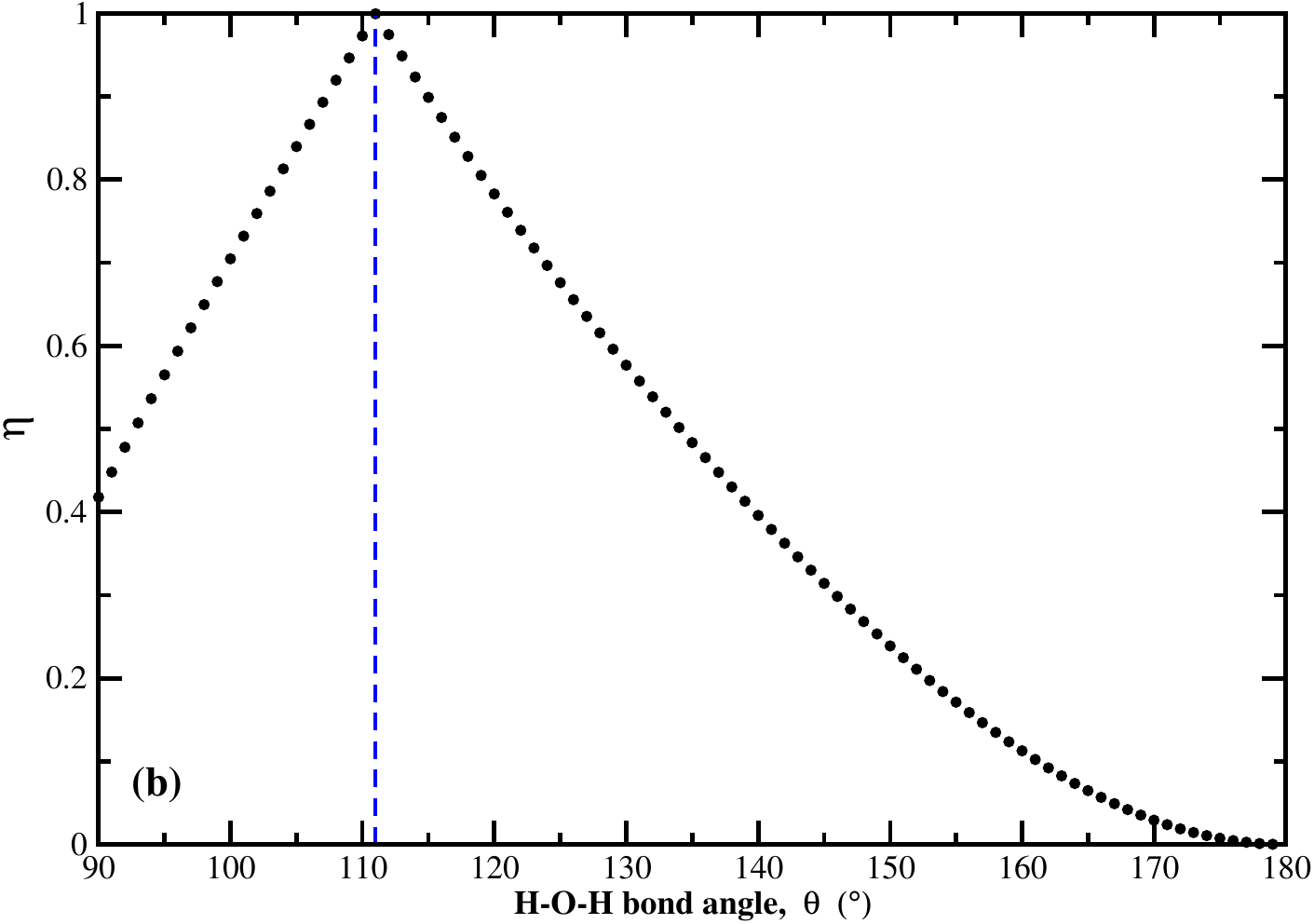}}

\caption{EFG at the oxygen in a water molecule as a function of the H–O–H bond angle, \(\theta\).
Panel (a) shows the principal EFG component \(V_{zz}\) (in a.u.), with solid black circles for \(V_{zz}\)
and open red circles for \(|V_{zz}|\). Panel (b) shows the corresponding asymmetry parameter \(\eta\).
%Calculations were performed using MOLCAS at the Hartree-Fock theory level with the ANO-L-VTZP basis set.
}
\label{fig:efg_geom}
\end{figure}

In Sec.~\ref{sec:symmetryefg} we discussed the symmetry properties of the EFG tensor and the nonphysical sensitivity of the sign of $V_{zz}$ to small perturbations when $\eta \approx 1$. Here we illustrate this effect using an isolated \ce{H2O} molecule as a representative molecular example. 
The equilibrium H--O--H angle in \ce{H2O} is $104.5^\circ$, while the linear configuration corresponds to $180^\circ$~\cite{Peng1996_water180}. To analyze the dependence of the EFG on molecular geometry, we varied the bending angle $\theta$ while keeping the O--H bond length fixed. Figure~\ref{fig:efg_geom} shows the principal component $V_{zz}$ and the asymmetry parameter $\eta$ at the oxygen nucleus as functions of $\theta$.
The magnitude $|V_{zz}|$ varies smoothly with geometry; however, near $\theta \approx 111^\circ$, where $\eta \to 1$, the sign of $V_{zz}$ changes discontinuously. This behavior reflects the ambiguity in the ordering of the principal components when $|V_{xx}|$ becomes small, and is therefore a consequence of the sorting convention in Eq.~\eqref{Eq:ViiSortingConvention}, rather than a physical effect.
For larger angles, $\eta$ decreases monotonically and approaches zero as $\theta \to 180^\circ$, where the molecule acquires cylindrical symmetry.

\begin{figure}[ht!]
    \centering
    \includegraphics[width=0.6\linewidth]{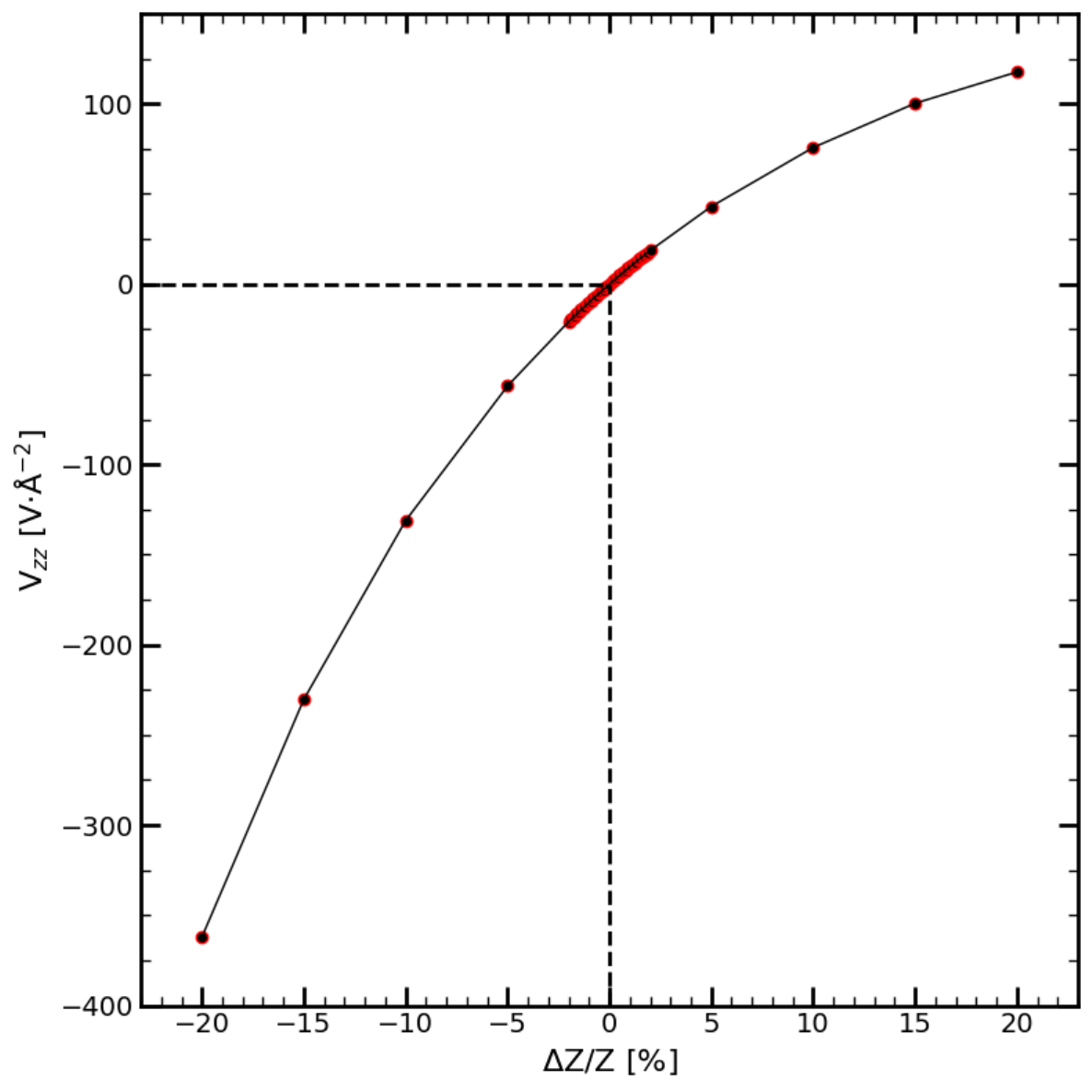}%\includegraphics[width=.5\linewidth]{dz:z2.png}
    % \caption{For the symmetrized cubic structure of \ce{LaCoO3} we  show influence of distortion of the Co–O bond lengths along the z-axis (a relative change of length) on the principal component \(V_{zz}\) of the electric field gradient tensor. In all cases the asymmetry parameter is null ($\eta = 0$). Calculations were performed at the density functional theory (DFT) level employing the PBE functional and the ANO-RCC-VQZP basis set. }

\caption{
 The effect of distortions in the Co--O bond lengths  along the $z$ axis (expressed as a relative change in bond length $Z$) on the principal component $V_{zz}$ of the electric-field–gradient tensor. This plot is for the symmetrized cubic structure of \ce{LaCoO3}. In all cases the asymmetry parameter vanishes ($\eta = 0$). 
 %The calculations were performed at the density-functional-theory level using the PBE functional and the ANO-RCC-VQZP basis set.
 }

    \label{fig:tet-dist}
\end{figure}

In Sec.~\ref{sec:EFG-clusters} we emphasized the importance of testing the robustness of embedded-cluster models. In particular, one expects the magnitude $|V_{zz}|$ at the central ion to vary continuously under small structural perturbations. As a test case, we consider \ce{LaCoO3}. In the cubic limit, both $V_{zz}$ and $\eta$ vanish by symmetry, corresponding to an ideal octahedral environment.
We introduce a controlled tetragonal distortion by varying the $(\text{Co--O})_z$ bond length while keeping the in-plane geometry fixed, which implies $V_{xx}=V_{yy}$ and hence $\eta=0$. The resulting $V_{zz}$ values are shown in Fig.~\ref{fig:tet-dist}.
For distortions up to $\sim 2\%$, $V_{zz}$ varies approximately linearly within $\pm 10\,$\efgunit, while larger distortions lead to a nonlinear response. The asymmetry between compression and elongation reflects the $1/r^3$ dependence of the EFG (Eq.~\eqref{Eq:EFG_def}). For this \ce{LaCoO3} test case, the calculated EFG responds smoothly and with the expected symmetry to small controlled distortions. This provides an internal consistency test of the embedded-cluster model, but does not by itself establish the absolute accuracy of the full deformation dependence for other materials.

\subsection{EFG sensitivity to computational parameters}\label{sec:active_space_sensitivity}

\begin{table}[h!]
%\centering
\caption{Active space size and EFG parameters for water molecule (one inactive orbital, 8 active electrons) and \ce{ThF8Ca12^{20+}} cluster (188 inactive orbitals, 6 active electrons). 
%CASSCF calculations were carried out using relativistic basis sets. The number of inactive orbitals and active electrons is reported in the table shown below for each system.
} \label{tab:active2}
%\apd{Add experimental results for water with error bars}
%\udeshika{The sign for H2O is not consistent with Table 1 (-181.3)}
%\begin{ruledtabular}
\begin{tabular}{crccrc}
System & Active space size  & Correlation & Dipole moment, & $V_{zz}$, & $\eta$ \\
  &   & Energy, a.u. &  Debye &  \efgunit &  \\
\midrule
 \ce{H2O} &HF \hspace{4pt}$\xrightarrow{} \hspace{9pt} $	4	&	0.000	&	1.98E+00	&	-181.3	&	0.79	\\
&+$2b_1$ anti-bonding \hspace{4pt}$\xrightarrow{} \hspace{9pt} $	5	&	0.022	&	1.92E+00	&	-179.2	&	0.78	\\
&+$4a_1$ anti-bonding \hspace{4pt}$\xrightarrow{} \hspace{9pt} $	6	&	0.054	&	1.83E+00	&	-171.0	&	0.78	\\
&diffuse orbitals begin \hspace{4pt}$\xrightarrow{} \hspace{9pt} $	7	&	0.095	&	1.78E+00	&	-163.6	&	0.81	\\
&	8	&	0.131	&	1.90E+00	&	-163.2	&	0.81	\\
&	9	&	0.140	&	1.86E+00	&	-165.4	&	0.78	\\
&	10	&	0.155	&	1.88E+00	&	-166.3	&	0.77	\\
&	11	&	0.167	&	1.88E+00	&	-164.9	&	0.78	\\
&	12	&	0.187	&	1.87E+00	&	-164.5	&	0.78	\\
&	13	&	0.208	&	1.89E+00	&	-164.6	&	0.77	\\
&	14	&	0.213	&	1.89E+00	&	-165.0	&	0.76	\\
%Experiment&       &           &               &   -165(2)        &   0.75(2)        \\
% &	15	&	0.219	&	1.88E+00	&	-166.9	&	0.77	\\
% &	16	&	0.223	&	1.86E+00	&	-167.2	&	0.76	\\
% &	17	&	0.227	&	1.84E+00	&	-167.2	&	0.77	\\
% &	18	&	0.231	&	1.85E+00	&	-167.4	&	0.77	\\
% &	19	&	0.234	&	1.85E+00	&	-167.4	&	0.77	\\
% &	20	&	0.239	&	1.82E+00	&	-167.4	&	0.77	\\
% &	21	&	0.243	&	1.83E+00	&	-167.6	&	0.77	\\
% &	22	&	0.248	&	1.82E+00	&	-167.6	&	0.77	\\
% &	23	&	0.252	&	1.82E+00	&	-167.7	&	0.77	\\
\midrule 
\ce{ThF8Ca12^{+20}}&HF \hspace{4pt}$\xrightarrow{} \hspace{9pt} $3 	&	0.000	&	9.34E-05	&	0.0	&	0.00	\\
&$6d$-subshell begins\hspace{7pt}$\xrightarrow{} \hspace{9pt} $4	&	0.007	&	9.35E-05	&	-122.3	&	0.00	\\
&5	&	0.017	&	9.37E-05	&	0.0	&	0.00	\\
&6	&	0.034	&	1.45E-04	&	-89.8	&	0.00	\\
&7	&	0.050	&	6.22E-02	&	-60.3	&	0.00	\\
&$6d$-subshell ends\hspace{7pt}$\xrightarrow{} \hspace{9pt}$8	&	0.070	&	5.42E-04	&	0.0	&	0.00	\\
&$7s$-subshell\hspace{5pt} $\xrightarrow{} \hspace{9pt} $9	&	0.073	&	3.43E-04	&	0.0	&	0.00	\\
&$5f$-subshell begins\hspace{7pt}$\xrightarrow{} \hspace{4pt} $10	&	0.076	&	3.38E-02	&	2.3	&	0.88	\\
&11	&	0.079	&	5.71E-02	&	-11.8	&	0.54	\\
&12	&	0.082	&	4.93E-02	&	15.4	&	0.90	\\
&13	&	0.086	&	4.67E-02	&	26.9	&	0.16	\\
&14	&	0.089	&	9.43E-05	&	6.7	&	0.00	\\
&15	&	0.092	&	9.69E-05	&	0.0	&	0.00	\\
&$5f$-subshell ends\hspace{7pt}$\xrightarrow{} \hspace{4pt} $16	&	0.097	&	9.50E-05	&	0.0	&	0.00	\\
\end{tabular}
%\end{ruledtabular}
\end{table}

In Sec.~\ref{sec:symmetryefg} we discussed the conditions that must be imposed on the choice of the active space to ensure that multiconfigurational CASSCF calculations preserve symmetry in EFG computations. While EFGs are generally weakly dependent on the choice of basis set (once a sufficient level of completeness is reached), they can be sensitive to the construction of the multiconfigurational wave function, where incomplete or unbalanced active spaces may break symmetry.

The selection of the active space is a critical step in any multiconfigurational treatment, as it defines the first-order interaction space in which the essential physics of the system is represented. The active space must include all symmetry-equivalent orbitals in order to permit their mutual rotations and mixing. In what follows, we explore several alternative active-space choices and demonstrate that, although the correlation energy may improve, symmetry breaking leads to nonphysical EFG results.

Because the choice of basis set has only a minor influence on the EFG parameters (provided that the basis is of sufficient quality, as discussed in the Supplementary Materials), we perform these tests using basis sets of at least TZP quality. Detailed information on the specific basis sets employed is given in the Supplementary Materials.

We begin with the \ce{H2O} molecule, correlating all electrons except the O~$1s$ core. The Hartree--Fock reference is constructed from four molecular orbitals: $2a_1$ (primarily O~$2s$), $b_2$ and $3a_1$ (bonding combinations of O~$2p$ and H~$1s$), and the lone-pair orbital.

The active space is then expanded incrementally by adding one orbital at a time, including the antibonding $4a_1$ and $2b_2$ orbitals, followed by diffuse orbitals with dominant O-centered $s$, $p$, and $d$ character. Table~\ref{tab:active2} shows the dependence of $V_{zz}$, $\eta$, and the dipole moment on the active-space construction. 

While the correlation energy increases monotonically with the size of the active space, the dipole moment and EFG parameters converge toward stable values. The largest change occurs upon inclusion of the first diffuse orbital.

We next consider the \ce{ThF8Ca12} cluster embedded in the \ce{CaF2} lattice, modeling a substitutional defect in which Th replaces Ca ($\mathrm{Th}_{\mathrm{Ca}}^{\ddot{\phantom{}}}$ in Kröger–Vink notation~\cite{kroger1974-book}). \thor{}-doped crystals are of interest for solid-state nuclear-clock platforms~\cite{Tiedau2024,Elwell2024}, and EFG-split \thor{} manifolds have been observed experimentally~\cite{Zhang2024-Th229Comb,Hiraki2025}.

In this system, the \ce{Th^{4+}} impurity has a closed-shell ground state, and the molecular orbitals are largely localized. The active space is therefore constructed from orbitals with predominantly atomic character, starting from the Th~$6p$ subshell (six electrons in three orbitals), while the remaining occupied orbitals are kept inactive.

The active space is systematically expanded up to 16 orbitals by including additional Th-centered $d$- and $f$-type orbitals. As shown in Table~\ref{tab:active2}, the correlation energy increases with the size of the active space, but the EFG parameters do not converge monotonically and may deviate from the symmetry-expected values $V_{zz}=0$ and $\eta=0$. These deviations are often correlated with changes in the dipole moment, indicating distortions in the charge distribution.

The origin of this behavior is the incomplete representation of symmetry-adapted orbital manifolds. For example, inclusion of a single Th~$6d$ orbital breaks cubic symmetry, whereas inclusion of the full $6d$ subshell (both $e_g$ and $t_{2g}$ components) restores it. A similar pattern is observed for the $5f$ orbitals, which in the $O_h$ point group decompose as $A_{2u} \oplus T_{1u} \oplus T_{2u}$: partial inclusion leads to symmetry breaking, while the full subshell recovers the correct symmetry.

These results demonstrate that, although increasing the active space improves correlation energy, it may simultaneously introduce symmetry breaking if orbital manifolds are not included in a balanced manner. Reliable EFG predictions therefore require symmetry-consistent active spaces, ideally including complete subshells. The dipole moment may serve as an auxiliary diagnostic, although it is less sensitive than $V_{zz}$ and $\eta$.

\subsection*{Conclusions}

We have examined selected factors controlling electric-field gradients
(EFGs) using representative molecular and solid-state examples, with emphasis
on multiconfigurational embedded-cluster calculations for solids. Different
quantum-chemical packages were shown to employ different sign conventions for
the principal component $V_{zz}$, which must be taken into account when
comparing computational results.

We further demonstrated that the sign of $V_{zz}$ may change
discontinuously when $\eta\approx1$, where the ordering of the principal
components becomes ambiguous. This behavior is a consequence of the
conventional ordering of the EFG tensor and does not correspond to a
physical change in the system.

For the crystalline examples considered here, the embedded-cluster
construction preserves the symmetry constraints on the EFG when both the
embedding and the multiconfigurational wave function are constructed
consistently. In particular, the choice of active space is crucial: incomplete
symmetry-equivalent orbital manifolds may lead to symmetry breaking and
spurious EFGs, whereas symmetry-consistent active spaces recover the expected
behavior. Once such numerical and symmetry-related deficiencies are removed,
physically meaningful geometry changes can produce substantially larger EFG
variations than further refinements of otherwise well-converged computational
details.

Overall, this work provides practical guidelines for reliable EFG
calculations and their interpretation in molecular and solid-state
systems, emphasizing consistent conventions, preservation of symmetry,
and balanced multiconfigurational descriptions.

\section{Acknowledgment}
AD thanks  Eric Hudson and Danny Rehn for discussions. VV thanks Dr.\ Adrian M. V. Br$\hat{a}$nzanic for performing Gaussian calculations. VV thanks 
Prof.\ Roland Lindh for a useful hint about implementation of EFG in MOLCAS. 
The work of AD and UP was supported in part by the U.S. NSF awards 
 PHY-2207546, PHY-2412869, and PHY-2513134.
% PHY-2412869 is with JW
VV and MK thank  Polish National Agency
for Academic Exchange under the Strategic Partnership Programme grant BNI/PST/2023/1/00013/U/00001, eSSENCE@LU 11:2, and grant WISE-WASP-02-05.  
This work used Bridges-2 at Pittsburgh Supercomputing Center through allocation PHY230110 from the Advanced Cyberinfrastructure Coordination Ecosystem: Services \& Support (ACCESS) program, which is supported by National Science Foundation grants \#2138259, \#2138286, \#2138307, \#2137603, and \#2138296 and computational resources provided by LUNARC (The Centre for Scientific and Technical Computing at Lund University) and NAISS 2025-22-655.

\nocite{POBTZVP} % cited only in the SI; mciteplus needs it registered before the bibliography
\bibliography{main}

\newpage

\section*{Supporting information}

\subsection*{Basis sets}

For the orbital-basis calculations, basis sets of at least triple-$\zeta$ quality were used whenever possible. We use a family of atomic natural orbital basis sets with relativistic corrections (ANO-RCC). In \textsc{MOLCAS}, the ANO basis set family is optimized for multiconfigurational calculations and allows for systematic improvement of basis-set quality. In other quantum-chemistry codes, where ANO-RCC is not available, compatible program-specific basis representations were used. The purpose of the calculations collected in Table~\ref{tab:efg_hf} is not to compare basis-set families, but to identify program-dependent EFG conventions. The basis-set test below shows that excessively small bases can alter the EFG substantially, whereas the dependence becomes small once a sufficiently flexible basis is reached. The explicitly tabulated molecular basis sets are given in Table~\ref{tab:molecular_bases}.

In embedded-cluster calculations, the primary goal is to describe the electronic structure of the central atom as accurately as possible. This is achieved by using the largest basis set for the central atom, while applying smaller basis sets to the surrounding coordination shells, so that the overall computational cost remains manageable with minimal impact on accuracy. The basis sets used for different atoms in the cluster calculations are summarized in Table~\ref{tab:cluster_bases}. 

\begin{table}[!ht]
    \centering
    \caption{Basis sets used for the molecular EFG calculations.}
    \label{tab:molecular_bases}
%    \begin{ruledtabular}
    \begin{tabular}{ccl}
                
                Method & Software & Basis \\ 
                \hline
         HF/DFT & MOLCAS & ANO-RCC-VQZP (O:5s4p3d2f1g, H:4s3p2d1f) \\
                    & ORCA &  Def2-TZVP\\        
                & Gaussian & aug-cc-pVTZ \\  
                \hline
        CASSCF/CASPT2 & MOLCAS& ANO-RCC-VQZP (O:5s4p3d2f1g, H:4s3p2d1f)\\
    \end{tabular}
%    \end{ruledtabular}
\end{table}

\begin{table}[!ht]
    \centering
    \caption{Basis sets used in calculation of EFG in embedded clusters. The boldfaced element represents the central atom of a given cluster. Atomic centers are listed from left to right, beginning with the central atoms, followed by those in the nearest coordination sphere.}\label{tab:cluster_bases}
%    \begin{ruledtabular}
    \begin{tabular}{cccccc}
        System&Cluster & \multicolumn{3}{c}{Atomic centers and corresponding bases}\\ \midrule
        \ce{CaF2}&\ce{\textbf{Ca}F8Ca12^{18+}} & \textbf{Ca} & F & Ca \\
                      &&ANO-RCC-VTZ   & ANO-RCC-VTZ  & ANO-XS-VDZ   \\ \midrule
        \ce{Th\colon CaF2}&\ce{\textbf{Th}F8Ca12^{20+}} & \textbf{Th} & F & Ca \\
                      &&ANO-RCC-VTZP     &ANO-RCC-VTZP    &ANO-XS-VDZP    \\ \midrule
        \ce{LaCoO3}&\ce{\textbf{Co}O6^{9-}} & \textbf{Co} & O &  \\
                      &&ANO-RCC-VQZP    &ANO-RCC-VQZP   &    \\
    \end{tabular}
%    \end{ruledtabular}
\end{table}
Periodic DFT calculations were performed with the \textsc{VASP} code, using a plane-wave basis set with a cutoff energy of 500~eV. Core electrons were treated with the projector augmented-wave (PAW) method.
The Hartree--Fock calculations for H$_2$O in \textsc{CRYSTAL} employed the POB-TZVP basis set~\cite{POBTZVP}, namely O: (10s6p1d)/[4s3p1d] and H: (5s1p)/[3s1p].

 % The funcionals were specified in the tables if it not the  other cases in my calculations it's PBE.

\subsection*{EFG sensitivity to the basis set}

We examined the dependence of EFG values on the choice of basis set (see Table~\ref{contraction}). Minimal basis sets, which include only valence-shell functions, yield qualitatively incorrect results, including even the sign of the EFG. In contrast, for basis sets of triple- to quadruple-$\zeta$ quality, the EFG values are well converged and show negligible variation.

\begin{table}[!ht]
\centering
\caption{Effect of \textbf{basis set size} on EFG eigenvalues and asymmetry parameter at the \textbf{HF} level for \ce{H2O}.
}
\label{contraction}
\begin{tabular}{lllccc}
\toprule 
Family & Label & Basis set contraction for O atom & Energy (Ha) & $V_{zz}$ (\efgunit) & $\eta$ \\
\midrule
ANO-RCC &MB& 2S1P   & -75.997 & 325.8 & 0.939 \\
        &VDZ& 3S2P   & -76.060 & -201.7 & 0.964 \\
        &VDZP& 3S2P1D & -76.108 & -185.3 & 0.913 \\
        &VTZP& 4S3P2D1F & -76.117 & -174.2 & 0.927 \\
        &VQZP& 5S4P3D2F1G & -76.118 & -177.6 & 0.918 \\
\midrule
ANO-L   &MB& 2S1P   & -75.941 & 329.7 & 0.935 \\
        &VDZ& 3S2P   & -76.005 & -203.9 & 0.937 \\
        &VDZP& 3S2P1D & -76.054 & -187.2 & 0.940 \\
        &VTZ& 4S3P2D & -76.062 & -177.4 & 0.945 \\
        &VTZP& 4S3P2D1F & -76.065 & -173.9 & 0.929 \\
        &VQZP& 5S4P3D2F & -76.066 & -177.5 & 0.939 \\
\midrule
Pople   & 6-31G   & 2S2P & -75.985 & -189.9 & 0.876 \\
        & 6-31G*  & 3S2P1D & -76.010 & -182.8 & 0.865 \\
        & 6-311G**& 4S3P1D & -76.046 & -193.9 & 0.924 \\ 
\midrule
Slater  & STO-3G  & 2S1P & -74.961 & 255.3 & 0.871 \\
\bottomrule
\end{tabular}

\end{table}

\subsection{Crystal structures used in this work}

\begin{table}[!ht]
\centering
\caption{Unit cell parameters for CaF$_2$.}
\begin{tabular}{lc}
\toprule \midrule
Formula & CaF$_2$ \\
Crystal system & Cubic \\
Space group & Fm$\bar{3}$m (No.\ 225) \\
$a$ (\AA) & 5.51605 \\ % Replace with your value
$b$ (\AA) & 5.51605 \\
$c$ (\AA) & 5.51605 \\
$\alpha = \beta = \gamma$ (°) & 90 \\
\midrule \bottomrule
\end{tabular}
\end{table}

\begin{table}[!ht]
\centering
\caption{Fractional atomic coordinates for CaF$_2$ in space group Fm$\bar{3}$m.}
\begin{tabular}{lccc}
\hline
Atom & $x$ & $y$ & $z$ \\
\hline
Ca (4a) & 0 & 0 & 0 \\
F (8c) & 0.25 & 0.25 & 0.25 \\
\hline
\end{tabular}
\end{table}

\begin{table}[h!]
\centering
\caption{Unit cell parameters for LaCoO$_3$.}
\begin{tabular}{lc}
\hline
Formula & LaCoO$_3$ \\
Crystal system & Rhombohedral \\
Space group & R$\bar{3}$c (No.\ 167) \\
$a$ (\AA) & 5.34477 \\ 
$b$ (\AA) & 5.34477 \\
$c$ (\AA) & 13.09195 \\ 
$\alpha = \beta = 90^\circ$, $\gamma = 120^\circ$ & \\
\hline
\end{tabular}

\end{table}

\begin{table}[h!]
\centering
\caption{Fractional atomic coordinates for LaCoO$_3$ in space group R$\bar{3}$c.}
\begin{tabular}{lcccc}
\hline
Atom & Wyckoff & $x$ & $y$ & $z$ \\
\hline
La & 6a & 0 & 0 & 0.25 \\
Co & 6b & 0 & 0 & 0 \\
O & 18e & 0.55265 & 0 & 0.25 \\ % 
\hline
\end{tabular}

\end{table}

%\section{EFG and selection of active space}

\newpage
\subsection*{Guide to computing electric field gradients in \textsc{MOLCAS}}

\textsc{MOLCAS}/\textsc{OpenMolcas} prints (when the \texttt{FLDG} keyword is used) the Cartesian components of the EFG tensor, from which the principal component $V_{zz}$ and the asymmetry parameter $\eta$ can be obtained by post-processing.

The EFG tensor $\mathbf{V}$ is real, symmetric, and traceless. Its principal values $(V_{xx}, V_{yy}, V_{zz})$ are defined in the principal-axis frame (PAF) with the conventional ordering $\lvert V_{zz}\rvert \ge \lvert V_{yy}\rvert \ge \lvert V_{xx}\rvert$ (see Sec.~\ref{Sec:EFGs-HQ:Conventions}). The asymmetry parameter is defined as
\begin{equation}
\eta = \frac{V_{xx}-V_{yy}}{V_{zz}}, \qquad 0 \le \eta \le 1.
\label{eq:eta}
\end{equation}

\paragraph*{Workflow.}
\begin{enumerate}
  \item Increase the \textsc{MOLCAS} print level either by setting \texttt{MOLCAS\_PRINT=4}, or by recompiling with \texttt{iPL=4} in \texttt{property\_util/prop.f}.
  \item Run \texttt{GATEWAY} with \texttt{FLDG=0} to print the Cartesian EFG tensor at each nucleus.
  \item For CASPT2 calculations, use the \texttt{DENSITY} keyword.
  \item Assemble the symmetric matrix
  \[
  \mathbf{V}=
  \begin{pmatrix}
  V_{xx} & V_{xy} & V_{xz} \\
  V_{xy} & V_{yy} & V_{yz} \\
  V_{xz} & V_{yz} & V_{zz}
  \end{pmatrix}
  \]
  from the output. Note that, in \textsc{MOLCAS}, the EFG tensor has the opposite sign to the convention used in Sec.~\ref{Sec:EFGs-HQ:Conventions}, so all components $V_{ij}$ must be sign-inverted.
  \item Diagonalize $\mathbf{V}$ to obtain eigenvalues and reorder them so that $\lvert V_{zz}\rvert \ge \lvert V_{yy}\rvert \ge \lvert V_{xx}\rvert$.
  \item Compute $\eta$ using Eq.~\eqref{eq:eta}.
\end{enumerate}

\paragraph*{Illustrative example.}

As an example, consider the bent H$_2$O molecule with an H--O--H angle of $\sim 104.5^\circ$. The $3\times 3$ EFG tensor is diagonalized to obtain its principal components, which are then ordered according to $\lvert V_{zz}\rvert \ge \lvert V_{yy}\rvert \ge \lvert V_{xx}\rvert$. The asymmetry parameter $\eta$ is subsequently evaluated using Eq.~\eqref{eq:eta}.

In atomic units (a.u.), the tensor reads
\[
\mathbf{V} =
\begin{pmatrix}
-1.866 & 0.000 & 0.000 \\
0.000 & 0.334 & -0.410 \\
0.000 & -0.410 & 1.542
\end{pmatrix},
\]
which yields the principal components
\[
V_{zz} = -1.866,\qquad V_{yy} = 1.667,\qquad V_{xx} = 0.199.
\]

The asymmetry parameter is
\[
\eta = \frac{V_{xx}-V_{yy}}{V_{zz}} \approx 0.767,
\]
which satisfies $0 \le \eta \le 1$.

For unit conversion, one atomic unit of EFG corresponds to 97.1736243~\efgunit, so $V_{zz}=-1.866$ a.u. corresponds to -181.33~\efgunit.
\end{document}